\documentclass[aps,prd,twocolumn,groupedaddress,amsfonts,amssymb,amsmath,showpacs,showkeys]{revtex4-1}
\usepackage{graphicx}

\usepackage{natbib}

\newcommand{\aap}{Astronomy and Astrophysics}

\newcommand{\mnras}{MNRAS}

\begin{document}


\title{A cosmological dust model with extended $f(\chi)$ gravity.}

\author{D.A. Carranza}
\email[Email address: ]{dgocarranza@gmail.com}
\author{S. Mendoza}
\email[Email address: ]{sergio@astro.unam.mx}
\author{L.A. Torres}
\email[Email address: ]{luisfciencias@gmail.com}
\affiliation{Instituto de Astronom\'{\i}a, Universidad Nacional
                 Aut\'onoma de M\'exico, AP 70-264, Distrito Federal 04510,
	         M\'exico \\
            }

\date{\today}

\begin{abstract}
  Introducing a fundamental constant of nature with dimensions of
acceleration into the theory of gravity makes it possible to extend
gravity in a very consistent manner.  At the non-relativistic level a
MOND-like theory with a modification in the force sector is obtained,
which is the limit of a very general metric relativistic theory of
gravity.  Since the mass and length scales involved in the dynamics of
the whole universe require small accelerations of the order of Milgrom's
acceleration constant \( \mathsf{a}_0 \), it turns out that the relativistic
theory of gravity can be used to explain the expansion of the universe.
In this work it is explained how to use that relativistic theory
of gravity in such a way that the overall large-scale dynamics of the
universe can be treated in a pure metric approach without the need to
introduce dark matter and/or dark energy components.
\end{abstract}


\maketitle

\section{Introduction}
\label{introduction}

  Current cosmological data is generally explained introducing two
unknown mysterious dark components, namely dark matter and dark energy.
These ad hoc hypothesis represent a big cosmological paradigm, since
they arise due to the fact that Einstein's field equations are forced
to remain unchanged under certain observed astrophysical and cosmological 
anomalies.

  A natural alternative scenario would be to see whether viable
cosmological solutions can be found if dark unknown entities are
assumed non-existent.  The price to pay with this assumption is that
the field equations of the theory of gravity need to be extended, and so
new Friedmann-like equations will arise.  The most natural approach to
extend gravity arises when a metric extension \( f(R) \) is introduced
into the theory \citep[see e.g.][and references therein]{capozziellobook}.

  In a series of recent articles,
\citet{mendoza11,bernal11,hernandez10,hernandez12a,bernal11a} have shown
how relevant the introduction of a new fundamental physical constant
\( \mathsf{a}_0 \approx 10^{-10} \textrm{m}/\textrm{s}^2 \) with dimensions of
acceleration is in excellent agreement with different phenomenology at
many different astrophysical mass and length dimensions, from solar-system
to extragalactic scales.  The introduction of the so called Milgrom's
acceleration constant \( \mathsf{a}_0 \) in a description of gravity means that
any gravitational field produced by a certain distribution of mass (and
hence energy) needs to incorporate the acceleration \( \mathsf{a}_0 \)
together with Newton's gravitational constant \( G \) and the speed of
light \( c \) in the description of gravity.  

 In this article we explore the consequences of the relativistic 
extended metric \( f(\chi) \) theory of gravity \citep{bernal11} applied to
the present expansion of the universe.  The article is organised as
follows.  Section~\ref{relativistic-extension} gives a brief summary of the
extended theory of gravity by \citet{bernal11}, generalising the Newnotian
description by \citet{mendoza11}.  Section~\ref{frt-connection}  
interconnects this extended relativistic
description of gravity with a metric description of gravity for which
the energy-momentum tensor appears in the gravitational field's action.
On section~\ref{cosmological-applications} we use the developed theory of
gravity for cosmological applications in a dust universe and see how it is
a coherent representation of gravity at cosmological scales.  Finally on
section~\ref{discussion}, we discuss the consequences of the developed
approach of gravity and some of the future developments of the theory.

\section{Relativistic metric extension}
\label{relativistic-extension}

  Finding a relativistic theory of gravity for which one
of its non-relativistic limits converges to MOND yields
usually strange assumptions and/or complicated ideas \citep[see
e.g.][]{mishra12,blanchet12,bekenstein04}.  A good first approach was
provided by a slight modification of Einstein's field equations by
\citet{sobouti06}, but the attempt is not complete.

  In order to find an elegant and simple theory of gravity with a
non-relativistic  weak-field limit MONDian solution, \citet{bernal11}
used a correct metric interpretation of Hilbert's action \( S_\text{f}
\) in such a way that:

\begin{equation}
   S_\text{f}  = - \frac{ c^3 }{ 16 \pi G L_M^2 } \int{ f(\chi) \sqrt{-g}
     \, \mathrm{d}^4x},
\label{eq010}
\end{equation}

\noindent which slightly differs from its traditional form (see
e.g. \cite{capozziellobook,sotiriou10,capozziello10a}) since the
dimensionless Ricci scalar:

\begin{equation}
  \chi := L_M^2 R,
\label{eq011}
\end{equation}

\noindent has been introduced.  In equation~\eqref{eq011}, \( R \) is  the
traditional Ricci scalar and \( L_M \) defines a length
fixed by the parameters of the theory.  For \( f(\chi)
= \chi \) the standard Einstein-Hilbert action is obtained.
The matter action has its usual form:

\begin{equation}
  S_\text{m} = - \frac{ 1 }{ 2 c } \int{ {\cal L}_\text{m} \, \sqrt{-g} \,
    \mathrm{d}^4x },
\label{eq012}
\end{equation}

\noindent with \( {\cal L}_\text{m} \) the matter Lagrangian density of the
system.

The null variations of the complete action, i.e. \( \delta
\left( S_\text{H} + S_\text{m} \right) = 0 \), yield the following
field equations:

\begin{equation}
  \begin{split}
    f'(\chi) \, \chi_{\mu\nu} - \frac{ 1 }{ 2 } f(\chi) g_{\mu\nu} - L_M^2 &
      \left( \nabla_\mu \nabla_\nu -g_{\mu\nu} \Delta \right) f'(\chi)
  				\\
    &= \frac{ 8 \pi G L_M^2 }{ c^4} T_{\mu\nu},
  \end{split}
\label{eq013}
\end{equation}

\noindent where the dimensionless Ricci tensor \( \chi_{\mu\nu} \) 
is given by:

\begin{equation}
  \chi_{\mu\nu} := L_M^2 R_{\mu\nu},
\label{eq014}
\end{equation}

\noindent and \( R _{\mu\nu} \) represents the standard Ricci tensor.
The Laplace-Beltrami operator has been written as \( \Delta :=
\nabla^\alpha \nabla_\alpha \) and the prime denotes derivative with
respect to its argument.  The energy-momentum tensor \( T_{\mu\nu} \)
is defined through the following standard relation: \( \delta S_\text{m}
= - \left( 1 / 2 c \right) T_{\alpha\beta} \, \delta g^{\alpha\beta} \).
In here and in what follows, we choose a (\(+,-,-,-\)) signature for
the metric \( g_{\mu\nu} \) and use Einstein's summation convention over
repeated indices.

  The trace of equation~\eqref{eq013} is:
\begin{equation}
  f'(\chi) \, \chi  - 2 f(\chi) + 3 L_M^2  \, \Delta  f'(\chi) = 
    \frac{ 8 \pi G L_M^2 }{ c^4} T,
\label{eq015}
\end{equation}

\noindent where \( T := T^\alpha_\alpha \).

  \citet{bernal11} showed that, for a point mass source and a spherically
symmetric space-time,  the choice 

\begin{equation}
  f(\chi) = \chi^{3/2},
\label{e01}
\end{equation}

\noindent has a MOND-like solution  at the weakest non-relativistic limit
of the theory (for which \( \chi \ll 1 \)) with:

\begin{equation}
  L_M = \zeta \, r_\text{g}^{1/2} l_M^{1/2}, \qquad \text{with} \qquad
    \zeta = \frac{2 \sqrt{2} }{ 9 },
\label{eqlm}
\end{equation}

\noindent where the ``mass-length'' scale \( l_M \)  and the gravitational
radius \( r_\text{g} \) are given by:

\begin{equation}
  l_M := \left( \frac{ G M }{ \mathsf{a}_0 } \right)^{1/2},  
    \qquad r_\text{g} := \frac{ G M }{ c^2 }.
\label{eq22}
\end{equation}

  In general terms, the function \( f(\chi) \) must satisfy the following
  limits:

\begin{equation}
   f(\chi) \rightarrow 
    \begin{cases}
      \chi^{3/2}, \quad \text{for } \chi \ll 1, \\
             \chi, \qquad \ \text{for } \chi \gg 1.  
    \end{cases}
\label{eq032}
\end{equation}

\noindent In other words, general relativity is recovered when \( \chi
\gg 1 \) in the strong field regime and the relativistic version of MOND
with \( \chi^{3/2} \) is recovered for the weak field regime of gravity
when \( \chi \ll 1 \). 

  The mass dependence of \( \chi \) and \( L_M \) means that Hilbert's
action~\eqref{eq010} is a function of the mass \( M \).  This is usually
not assumed, since that action is thought to be purely a function of the
geometry of space-time due to the presence of mass and energy sources.
However, it was Sobouti \cite{sobouti06} who first encountered
this peculiarity in the  Hilbert action when dealing with a metric
generalisation of MOND.  Following the remarks by \citet{sobouti06} and
\citet{mendoza07} one should not be surprised if some of the commonly
accepted notions, even at the fundamental level of the action, require
generalisations and re-thinking.  An extended metric theory of gravity
goes beyond the traditional general relativity ideas and in this way,
we need to change our standard view of its fundamental principles.

\section{$F(R,T)$ connection}
\label{frt-connection}

  For the description of gravity shown in
section~\ref{relativistic-extension} it follows that an adequate way of
writing up the gravitational field's action is given by:

\begin{equation}
   S_\text{f}  = - \frac{ c^3 }{ 16 \pi G  } \int{ \frac{ f(\chi) }{ L_M^2
     } \sqrt{-g} \, \mathrm{d}^4x}.
\label{eq040}
\end{equation}

\noindent The function \( L_M \) is a function of the mass of the system
and in general terms it is a function of the space-time coordinates.
For the particular case of a spherically symmetric space-time it coincides
with the mass of the central object generating the gravitational field as
expressed in equations~\eqref{eqlm} and~\eqref{eq22}.  Generally speaking
what the meaning of \( M \) would be for a particular distribution
of mass and energy needs further development, beyond the scope of
this work.  Nevertheless one expects that for a systems with high
degree of symmetry (such as as a spherically symmetric space-time or
a Friedmann-Lema\^{\i}tre-Robertson-Walker one), the function \(
M \) would be given by the standard mass-energy relation \citep[see
e.g.][]{MTW}:

\begin{equation}
  M := \frac{ 4 \pi }{ c^2 } \int{ T \, r^2  \, \mathrm{d}r},
\label{eq041}
\end{equation}

\noindent where \( r \) is the ``radial'' coordinate.

  The field equations produced by the null variations of the addition of
the field's action \( S_\text{f} + S_\text{m} \) can be constructed in the
following form.  \citet{harko11} have built an \( F(R,T) \) theory of
gravity, so making the natural identification:

\begin{equation}
  F(R,T) := \frac{ f(\chi)  }{ L_M^2 },
\label{eq03}
\end{equation}

\noindent it is possible to use all their results for our particular case
expressed in equation~\eqref{eq03}.
For example,  the null variations of the complete action 
\( S_\text{f} + S_\text{m} \) for the particular case of
equation~\eqref{eq03} yield the following field equations \citep{harko11}:

\begin{equation}
  \begin{split}
    \left( \frac{ f_R }{ L_M^2 } \right) \, R_{\mu\nu} -& \frac{ 1 }{
      2 L_M^2 }  f \, g_{\mu\nu} + \bigg[ g_{\mu\nu} \Delta - \Delta_\mu
      \Delta_\nu \bigg] \left( \frac{ f_R }{ L_M^2 } \right) =
     							\\ 
    &\frac{ 8 \pi G }{ c^4 } T_{\mu\nu} - \left( \frac{ f }{ L_M^2 }
      \right)_T \bigg[ T_{\mu\nu} + \Theta_{\mu\nu} \bigg],
  \end{split}
\label{eq07}
\end{equation}

\noindent with a trace:

\begin{equation}
  \frac{ f_R \, R }{  L_M^2 } - \frac{ 2 f }{ L_M^2 } + 3 \Delta \left(
    \frac{ f_R }{ L_M^2 } \right) = \frac{ 8 \pi G }{ c^4 } T - \left(
    \frac{ f }{ L_M^2 } \right)_T \bigg[ T + \Theta \bigg],
\label{eq08}
\end{equation}

\noindent where the subscripts \( R \) and \( T \) stand for the partial
derivatives with respect to those quantities, i.e.

\begin{equation}
  \bigg( \ \ \bigg)_R := \frac{ \partial }{ \partial R }, \qquad
    \text{and} \qquad
  \bigg( \ \ \bigg)_T := \frac{ \partial }{ \partial T}.
\label{eq08a}
\end{equation}

\noindent The tensor \( \Theta_{\mu\nu} \) is such that 
\( \Theta_{\mu\nu} \delta g^{\mu\nu} := g^{\alpha\beta} \delta
T_{\alpha\beta} \)  and for the case of an ideal fluid it can be written as
\citep{harko11}:

\begin{equation}
  \Theta_{\mu\nu} = - 2 T_{\mu\nu} - p g_{\mu\nu}.
\label{eq09}
\end{equation}

  Note that equation~\eqref{eq07} or~\eqref{eq08} converges to the
  field~\eqref{eq013}
and trace~\eqref{eq015} relations as discussed in
section~\ref{relativistic-extension} when one considers a
point mass generating the gravitational field, i.e. when \( L_M =
\text{const.} \) and so \( \partial / \partial R = L_M^2 \partial / \partial
\chi \).

  In general terms, the \( F(R,T) \) theory described by \citet{harko11}
produces non-geodesic motion of test particles since:

\begin{equation}
  \begin{split}
  \nabla ^{\mu } & T_{\mu \nu }
     = \left( \frac{ f }{ L_M^2}\right)_{T}
      \left\{ \frac{ 8\pi G }{ c^4 } - \left( \frac{ f }{ L_M^2
      }\right)_{T} \right\}^{-1} \times \\
    & \left[ \left( T_{\mu \nu }+\Theta
      _{\mu \nu }\right) \nabla^{\mu } \ln \left( \frac{f}{L_M^2}
      \right)_{T}\left( R,T\right) + \nabla ^{\mu }\Theta _{\mu \nu
      }\right] \neq 0,
  \end{split}
\label{eq10}
\end{equation}

\noindent and as such the geodesic equation has a force term:

\begin{equation}
  \frac{d^{2}x^{\mu }}{ds^{2}}+\Gamma _{\nu \lambda }^{\mu }u^{\nu
    }u^{\lambda }=\lambda^{\mu },  
\label{eqmot}
\end{equation}

\noindent where the four-force 

\begin{equation}
  \begin{split}
    \lambda^{\mu }: =& \frac{ 8 \pi G }{ c^4 } \left( \rho c^2 + p
      \right)^{-1} \left[ \frac{ 8 \pi G }{ c^4 } 
      + \left( \frac{ f }{ L_M^2 } \right)_T \right]^{-1} \times \\
    & \left( g^{\mu \nu }
      - u^{\mu } u^{\nu } \right) \nabla _{\nu }p,
  \end{split}
\end{equation}

\noindent is perpendicular to the four velocity \( \mathrm{d}x^\alpha /
\mathrm{d} s \).  As explained by \citet{harko11}, the motion of test
particles is geodesic, i.e. \( \lambda^\mu = 0 \) and/or \( \nabla^\alpha
T_{\alpha\beta} = 0 \),  (i) for the case of a pressureless \( p = 0 \)
(dust) fluid and (ii) for the cases in which \( F_T(R,T) = 0 \).

  In what follows we will see how all the previous ideas can be applied to
a FLRW dust universe and so, the divergence of the energy momentum tensor
in equation~\eqref{eq10} is null.  It is worth noting that this condition
on the energy-momentum tensor for many applications needs to be zero,
including applications to the universe at any epoch.  In other words, 
one should impose the null divergence of the energy momentum tensor for any
real physical system.

\section{Cosmological applications}
\label{cosmological-applications}

  For an isotropic Friedmann-Lema\^{\i}tre-Robertson-Walker (FLRW) universe,
the interval \( \mathrm{d} s \) is given by 
\citep[see e.g.][]{galaxy-formation}:

\begin{equation}
  \mathrm{d}s^2 = c^2 \mathrm{d}t^2 - a^2(t) \left\{ \frac{ 
    \mathrm{d}r^2 }{ 1 - \kappa r^2 } + r^2 \mathrm{d}\Omega^2 \right\},
\label{eq11}
\end{equation}

\noindent where \( a(t) \) is the scale factor of the universe normalised
to unity, i.e. \( a(t_0) = 1 \), at the present epoch \( t_0 \), and 
the angular displacement \( \mathrm{d} \Omega^2 := \mathrm{d} 
\theta^2 + \sin^2\theta \, \mathrm{d} \varphi^2 \) for the polar 
\( \mathrm{d} \theta \) and azimuthal \( \mathrm{d} \varphi \) 
angular displacements with a comoving coordinate distance \( r \).  
From now on, we assume a null space curvature \( \kappa = 0 \) at the present
epoch in accordance with observations and deal with the expansion of the
universe dictated by the field equations~\eqref{eq07}, avoiding any form
of dark unknown component.  Since we are interested on the compatibility
of this cosmological model with SNIa observations, in what follows we
assume a dust \( p = 0 \) model for which the covariant divergence
of the energy-momentum tensor vanishes, and so as discussed in
section~\ref{frt-connection} the trajectories of test particles are
geodesic, i.e. the divergence of the energy-momentum tensor is zero.
As mentioned at the end of the previous section, note also that this
condition has to be satisfied for any equation of state and not only for
a dust universe.

  For simplicity, let us consider that the function \( f(\chi) \) obeys a 
power-law relation:

\begin{equation}
  f(\chi) = \chi^b,
\label{fchi-power}
\end{equation}

\noindent with a constant exponent \( b \).  Let us now rewrite the
field equations~\eqref{eq07} inspired by the approach first introduced
by \citet{capozziello02} (see also \citet{capozziellobook}) as follows:

\begin{gather}
    G_{\mu\nu} = \frac{ 8 \pi G }{ c^4 } \left\{ \left( 1 + \frac{ c^4 }{ 
      8 \pi G } F_T \right) \frac{ T_{\mu\nu} }{  F_R } +
      T_{\mu\nu}^\text{curv} \right\},
					\label{eq12} \\
\intertext{where the Einstein tensor is given by its usual form:}
  G_{\mu\nu}:= R_{\mu\nu} - \frac{1}{2} R g_{\mu\nu}.
  					\label{eq13} \\
\intertext{and}
  \begin{split}
    T_{\mu\nu}^\text{curv} :=  \frac{ c^4 }{ 8\pi G F_R } 
      &\left[  \left(\frac{1}{2} \left( F - R F_R \right)
            - \Delta F_R \right) g_{\mu\nu} \right.  + \\
	    & \nabla_{\mu}\nabla_{\nu}F_R \bigg],
  \end{split}
					\label{eq14}
\end{gather} 

\noindent represents the ``\emph{energy-momentum}'' curvature tensor.  Since
\( T_{00} = \rho c^2 \),  then it will be useful the identification \(
T_{00} := \rho_\text{curv} c^2 \).  With this last definition and using the
fact that the Laplace-Beltrami operator applied to a scalar field \( \psi \) is
given by  \citep[see e.g.][]{daufields}:

\begin{equation}
  \Delta \psi = \frac{ 1 }{ \sqrt{-g} } \partial_\mu \left( 
    \sqrt{ -g } \, g^{\mu\nu} \partial_\nu \psi \right),
\label{laplace-beltrami}
\end{equation}

\noindent then 

\begin{equation}
  \rho_{\text{curv}} = \frac{ c^2 }{ 8\pi G F_R } \left[ \frac{ 1 }{ 2 }
    \left( R F_R -F \right) - \frac{ 3H }{ c^2 } \frac{ \mathrm{d} F_R 
    }{ \mathrm{d} t }\right],
\label{eq15}
\end{equation}

\noindent where \( H := \dot{a}(t)/ a(t) \) represents Hubble's constant.

  With the above definitions and using the \( 00 \) component of the
field's equations~\eqref{eq12} and the relation \citep[cf.][]{dalarsson}

\begin{equation}
  R = - \frac{ 6 }{ c^2 } \left[ \frac{ \ddot{a} }{ a } +
    \left( \frac{ \dot{a} }{ a } \right)^2 + \frac{ \kappa c^2 }{ a^2 }
    \right],
\label{eq16}
\end{equation}

\noindent between Ricci's scalar and the derivatives of the scale factor
for a FLRW universe, then the dynamical  Friedman's-like equation for a
dust flat universe is:

\begin{equation}
  H^2 = \frac{ 8\pi G }{ 3 } \left[ \left( 1 + \frac{ c^4 }{ 8\pi G }
    F_T \right) \frac{ \rho }{ F_R } + \rho_{ \text{curv} } \right].
\label{friedmann}
\end{equation}

  The energy conservation equation is given by the null divergence
of the energy-momentum tensor:

\begin{equation}
  \left( \frac{ 8\pi G }{ c^4 } + F_T \right) \left( \dot{\rho} + 
    3 H \rho \right) = - \rho \frac{\mathrm{d}F_T}{\mathrm{d}t} = 0,
\label{tempocontinuity}
\end{equation}

\noindent and so

\begin{equation}
  \dot{\rho} + 3 H \rho = 0.
\label{continuity}
\end{equation}

  For completeness, we write down the correspondent generalisation of 
Raychadhuri's equation for a dust flat universe:

\begin{equation}
  2 \frac{ \ddot{a} }{a} + H^2 = - \frac{ 8 \pi G p_\text{curv} }{c^2},
\label{raychadhuri}
\end{equation}

\noindent where the ``curvature-pressure'' 

\begin{equation}
  p_\text{curv} := \omega c^2 \rho_\text{curv},
\label{pcurv}
\end{equation}
  
\noindent and 

\begin{equation}
  w =  \frac{ c^2 \left( F - R F_R \right) / 2 + 
    \mathrm{d}^2 F_R / \mathrm{d} t^2  +  3 H 
    \mathrm{d} F_R / \mathrm{d} t }{ c^2  \left( R F_R 
    - F \right) / 2 -  3 H  \mathrm{ d } F_R /
    \mathrm{d} t }. 
\label{omega}
\end{equation}

   On the other hand, note that the mass \( M \) that appears on the
length \( L_M \) must be the causally connected mass at a certain cosmic
time \( t \), since particles beyond Hubble's (or particle) horizon with
respect to a given fundamental observer do not have any gravitational
influence on him.  In other words:

\begin{equation}
  M = \int_{0}^{r_\text{H}} \rho \, r^2 \, \mathrm{d} r 
    = \frac{4}{3}\pi \rho \frac{ c^3 }{ H^3 },
\label{mass}
\end{equation}

\noindent where 

\begin{equation}
  r_\text{H} := \frac{ c }{ H(t) },
\label{hubble-radius}
\end{equation}

\noindent is the Hubble radius or the distance of causal contact at a
particular cosmic epoch \citep{galaxy-formation}.  With such an election
for the mass \( M(t) \), it follows that at any fixed cosmic epoch the
field's action~\eqref{eq040} is the same for all fundamental observers.
Using equation~\eqref{hubble-radius}, the length~\eqref{eqlm} can be
written as:

\begin{equation}
  L_M = \zeta \frac{ \left( \frac{4}{3} \pi c^3 G \right)^{3/4} }{ 
    c \, \mathsf{a}_0^{1/4} } \frac{ \rho^{3/4} }{ H^{9/4} }, 
\label{eq19}
\end{equation}

\noindent and so, by using relation~\eqref{fchi-power} then:

\begin{gather}
  \frac{ \mathrm{d}F_R }{ \mathrm{d} t } = b (b-1) R^{b-1} 
    L_M^{2(b-1)} H \left[\frac{ j - q - 2 }{ 1 - q }+ \frac{ 3 }{ 2 }
    \left( \beta + \frac{ 3 }{ \alpha } \right) \right],
    					\label{eq20}\\
  \frac{ \mathrm{d}F_T }{ \mathrm{d}t } = \frac{ 3 }{ 2 }( b - 1 ) 
    \frac{ R^b L_M^{ 2b-2 } }{ \rho c^2 },
\label{derivadas_frft}
\end{gather}

\noindent where

\begin{equation}
  q(t):= -\frac{ 1 }{ a }\frac{ \mathrm{d}^2 a }{ \mathrm{d} t^2 } H^{-2}, 
    \quad \text{ and } \quad 
    j := \frac{ 1 }{ a } \frac{ \mathrm{d}^3 a }{ \mathrm{d} t^3 } H^{-3},
\label{j}
\end{equation}

\noindent are the deceleration parameter and the jerk respectively.

  As it is usually done, let us find power law solutions satisfying 

\begin{equation}
  a(t) = a(t_0) \left( \frac{ t }{ t_0 } \right)^{\alpha}, \qquad 
  \rho(t) = \rho_0 \left( \frac{ a }{ a(t_0) } \right)^{\beta},
\label{power-laws}
\end{equation}

\noindent for the unknown indices \( \alpha \) and \( \beta \).  With
these and the value of \( L_M \) from equation~\eqref{eq19}, the curvature
density~\eqref{eq15} is

\begin{equation}
  \rho_\text{curv} = \frac{ 3 H^2 }{ 8 \pi G } \left( b - 1 \right) 
    \left[ \left( 1 - q \right) - \frac{ j - q - 2}{ 1 - q } -
    \frac{ 3 }{ 2 } \left( \beta + \frac{ 3 }{ \alpha } \right) 
    \right].
\label{dens_curv1}
\end{equation}

\noindent Substitution of the previous relations on Friedmann's
equation~\eqref{friedmann} it follows that:

\begin{equation}
  H^2 = \frac{ 8 \pi G \rho }{ 3 \, Z \, F_R },
\label{eq21}
\end{equation}

\noindent where

\begin{equation}
  Z := 1 + \left( b - 1 \right) \left[ \frac{ j - q - 2 }{ 1 - q } - 
    \frac{ 4 \left( 1 - q \right) }{ b } + \frac{ 3 }{ 2 } \left(
    \beta + \frac{ 3 }{ \alpha } \right) \right],
\label{Z}
\end{equation}

\noindent is a dimensionless function.

  An important result can be obtained evaluating equation~\eqref{eq21}
at the present epoch, yielding:

\begin{equation}
  \mathsf{a}_0 = \left[ \frac{ 9 }{ 4 } \zeta^4 \left( 1 - q_0 \right)^2
    \left( b Z_0 \right)^{ 2 / \left( b - 1 \right) }
    \left( \Omega_\text{matt}^{(0)} \right)^{ \left( 3 b - 5 \right) /
    \left( b - 1 \right) } \right] c \, H_0,
\label{a0ch0}
\end{equation}

\noindent where the density parameter \( \Omega_\text{matt}^{ (0) } \) 
at the present epoch has been defined by it's usual relation:

\begin{equation}
  \Omega_\text{matt}^{ (0) } := \frac{ 3 H^2 \rho }{ 8 \pi G }.
\label{omegazero}
\end{equation}

  In other words, the value of Milgrom's acceleration constant \(
\mathsf{a}_0 \)  at the current cosmic epoch is such that 

\begin{equation}
  \mathsf{a}_0 \approx c \times H_0.
\label{eq23}
\end{equation}

\noindent The numerical coincidence between the value of Milgrom's
acceleration constant \( \mathsf{a}_0 \) and the  multiplication of the
speed of light \( c \) by the current value of Hubble's constant \(
H_0 \) has been noted since the early development of MOND \citep[see
e.g.][and references therein]{famaey11}.  Note that equation~\eqref{eq23}
means that this coincidence relation occurs at approximately the present
cosmic epoch in complete agreement with the results by \citet{bernal11a}
where it is shown that \( \mathsf{a}_0 \) shows no cosmological evolution
and hence it can be postulated as a fundamental constant of nature.

  For the power law~\eqref{fchi-power} and the
assumptions made above, it follows that the energy conservation
equation~\eqref{continuity} is given by:

\begin{equation}
  \left( \dot{\rho} + 3 H \rho \right) + \frac{ c^2 }{ 8 \pi G }
    \left( A \frac{ \dot{\rho} }{ \rho} +B \, H \right) R^b \, 
    L_M^{2(b-1)}=0,
\label{continuity-simplified}
\end{equation}

\noindent where:

\begin{gather*}
  A := \frac{ 9 }{ 4 }\left( b - 1 \right)^2,
  				\\
  B := \frac{ 9 }{ 2 } \frac{ b - 1 }{ b } + \frac{ 27 }{ 4 } \frac{
    \left( b - 1 \right)^2 }{ \alpha } + \frac{ 3 }{ 2 } 
    \frac{ b \left( b - 1 \right) \left( j - q - 2 \right) }{ 1 - q }.
\end{gather*}

\noindent Direct substitution of the density power law~\eqref{power-laws} into
relation~\eqref{continuity-simplified} gives a constraint equation between \(
\alpha \), \( \beta \) and \( b \):

\begin{equation}
  \beta = \frac{ 1 }{ \alpha } \left( \frac{ 9 - 5 b }{ 3 b - 5 }\right).
\label{constraint}
\end{equation}

\begin{figure}
\begin{center}
  \includegraphics[scale=0.70]{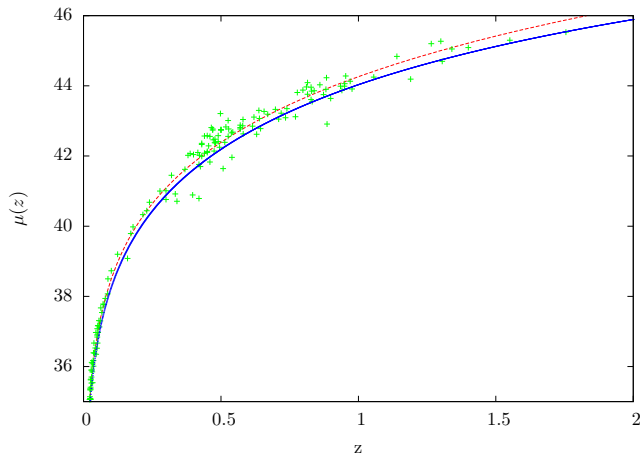}
  \caption[Best fit to SNIa data]{\label{fig02}
	  Redshift magnitude plot for SNIa showing the  distance modulus
	  \( \mu \) as a function of the redshift \( z \) for SNIa as
	  presented by \citet{riess04}.  The dotted red line shows the best
	  fit to the data with the \( f(\chi) \) gravity theory applied to
	  a flat dust FLRW universe (see text) with no dark components.	
	  The continuous blue line represents the best fit according to the
	  standard concordance dust \( \Lambda \)CDM model.
          }
\end{center}
\end{figure}

  Let us now proceed to fix the so far unknown parameters of the theory \(
\alpha \), \( \beta \) and \( b \).   To do so, we need reliable
observational data and as such, we use the redshift-magnitude SNIa data 
obtained by \citet{riess04} and the following well known 
standard cosmological relations~\citep[see e.g.][]{galaxy-formation}:

\begin{gather}
  1 + z = a(t_0) / a(t),
  				\label{eq24} \\
  \mu\left(z\right) = 5 \log_{10}\left[ H_0 \, d_L \left( z \right )\right] - 
    5 \log_{10} h + 42.38,
  				\label{eq25} \\
  d_\text{L}\left(z\right) = \left( 1 + z \right) \int_{0}^{z} 
    \frac{ c }{ H \left( z \right)}\,\ \mathrm{d}z,
 				\label{eq26}
\end{gather}

\noindent for the cosmological redshift \( z \), the distance modulus \(
\mu \), the Luminosity distance \( d_L \) and where the normalised Hubble
constant \( h \) at the present epoch is given by \( h := H_0 / \left(
100 \, \mathrm{km} \, \mathrm{ s }^{-1}  / \, \mathrm{Mpc} \right) \).
Also, from equation~\eqref{power-laws} it follows that

\begin{equation}
  H(a) = H_0\left(\frac{a}{a(t_0)}\right)^{-1/\alpha} = H_0(1+z)^{1/\alpha},
\label{solution_H}
\end{equation}
  
\noindent and the substitution of this into equation~\eqref{eq26} gives the
distance modulus \( d_L \) as a function of the redshift \( z \).  This
means that the redshift magnitude relation~\eqref{eq25} is a function that
depends on the values of the current Hubble constant \( H_0 \) and the
value of \( \alpha \).  Figure~\ref{fig02} shows the best fit to the
redshift magnitude relation of SNIa observed by \citet{riess04}, 
yielding \( \alpha=1.359 \pm 0.139 \) and  \( h = 0.64 \pm 0.009 \).  The
best fit presented on the figure was obtained using the Marquardt-Levenberg
fit provided by gnuplot (http://www.gnuplot.info) for non-linear functions.
This value does not conclude the whole description of the problem,
since \( \beta \) and \( b \) are still unknown but according to the
constraint equation~\eqref{constraint} only one of them is needed in order
to know the other once \( \alpha \) is known.

  The parameter \( \beta \) can be found from the conservation of mass
equation~\eqref{continuity}, which yields \( \rho \propto a^{-3} \)
and so \( beta = 3 \), which exactly coincides with the results of
standard cosmology when dealing with a dust FLRW universe \citep[see
e.g.][]{galaxy-formation}.  Using this value of \( \beta \) and the one
already found for \( \alpha \), it follows that \( b = 1.57 \pm 0.56
\) which is within the expected value of \( b = 3/2 \)  discussed in
section~\ref{relativistic-extension}.

  For completeness, we write down a few of the cosmographycal parameters
obtained by this \( f(\chi) \) gravity applied to the universe:

\begin{equation}
   \begin{gathered}
   h = 0.64 \pm 0.009,  \quad q_0 = -0.2642 \pm 0.075, \\
   j_0 = -0.1246 \pm 0.004.
   \end{gathered}
\label{eq27}
\end{equation}


\section{Discussion}
\label{discussion}

  The obtained value \( b \approx 3/2 \) is a completely
expected result due to the following arguments.  As explained by
\citet{mendoza11}, a gravitational system for which its
characteristic size \( r \) is such that \( x := l_M / r \lesssim 1 \)
is in the MONDian regime of gravity.  For the case of the universe, \( x
\sim \text{a few} \), and as such if not totally in the MONDian regime
of gravity, then it is far away from the regime of Newtonian gravity.
The relativistic version of this means that the universe is close to
the regime for which \( f( \chi ) = \chi^{3/2} \) and so \( b = 3/2 \).
This is a very important result since, seen in this way, the accelerated
expansion of the universe is due to an extended gravity theory deviating
from general relativity.  It is quite interesting to note that the
function \( f(\chi) = \chi^{3/2} \) which at its non-relativistic
limit is capable of predicting the correct dynamical behaviour of many
astrophysical phenomena, is also able to explain the behaviour of the
current accelerated expansion of the universe.

  Seen in this way, the behaviour of gravity towards the past (for
sufficiently large redshifts \( z \)) will differ from \( f(\chi) =
\chi^{3/2} \) and eventually converge to \( f(\chi) = \chi \), i.e. the
gravitational regime of gravity is general relativity for sufficiently
large redshifts. A very detailed investigation into this needs to be done
at different levels in order to be coherent with many different cosmological
observations \citep[see e.g.][]{longair11}.  

  It is quite remarkable that a metric extended theory of gravity is able
to reproduce phenomena from mass and length scales associated to 
the solar system up to cosmological scales.  Many other cosmological
applications of the theory will be addressed elsewhere.

\section{Acknowledgements}
\label{acknowledgements}

  This work was supported by two DGAPA-UNAM grants (PAPIIT IN116210-3
and IN111513-3).  DAC, SM and LAT thank support granted by  CONACyT: 48014, 
26344 and 221045.

\bibliographystyle{aipauth4-1}
\bibliography{a0isH0}

\end{document}